\documentclass[traditabstract]{aa} 
\usepackage[dvips]{graphicx}          

\newcommand\dnu{\Delta\nu}
\newcommand\deltanunu{\Delta\nu(\nu)}
\newcommand\dnumoy{\langle\Delta\nu\rangle}
\newcommand\numax{\nu\ind{max}}
\newcommand\nuenv{\delta\nu\ind{env}}
\newcommand\FWHM{\delta\nu\ind{env}}

\newcommand{\ind}[1]{_{\mathrm{#1}}}

\newcommand\diff{\, \mathrm{d}}

\begin{document}
   \title{An automated pipeline for asteroseismology based on the autocorrelation of stellar time series}


   \author{B. Mosser\inst{1}
               \and
           T. Appourchaux\inst{2}
          }

   \institute{LESIA, CNRS, Universit\'e Pierre et Marie Curie, Universit\'e Denis Diderot, Observatoire de Paris, 92195 Meudon cedex, France\\
        \email{benoit.mosser@obspm.fr}
\and
Institut d'Astrophysique Spatiale, UMR8617, Universit\'e Paris XI, B\^atiment 121, 91405 Orsay Cedex, France}

  \abstract
{The autocorrelation of an asteroseismic time series has been identified as a powerful tool capable of providing measurements of the large frequency separations. The performance of this method has been assessed and quantified by Mosser \& Appourchaux (2009). We propose now an automated pipeline based on it and describe its performance.
}

\maketitle

\section{Introduction}

The mean large separation in a seismic spectrum is a key parameter in asteroseismology. Observationally, identifying the large separation proves the detection of solar-like oscillation.
It was the case with the first determination of the exact large separation of Procyon (\cite{1998A&A...340..457M}) even if the energy excess of the oscillation was not identified, or of the CoRoT target HD 175726 for which no mode identification has been possible (\cite{mosser2009}). The measurement of the large separation opens then the determination of other parameters and the possibility of a large scientific output (stellar modeling, mass and radius measurement...).

With the event of large set of data produced by CoRoT and Kepler, it appeared necessary to develop efficient tools able to determine the parameters that describe an oscillation spectrum (\cite{2009AIPC.1170..540M, 2009arXiv0906.5002H, 2009arXiv0910.2764H}). All these tools search first in the power spectrum for an energy excess due to the solar-like oscillations. Then, when this excess has been identified, the signature of a solar-like oscillation spectrum is searched with different methods, all dealing with the detection of a regular pattern \`a la Tassoul (\cite{1980ApJS...43..469T}).

We have chosen a different way to perform the analysis of seismic time series. Instead of searching for the regularity of the oscillation pattern in the Fourier spectrum, we search for its physical cause. The comb structure of the solar-like oscillation pattern being due to the fact that all pressure modes propagate throughout the stellar diameter at the same phase velocity, we search for this signature in the autocorrelation of the time series. The autocorrelation correlates in fact any wave with itself after a double travel across the stellar diameter. Instead of computing the autocorrelation, we deal with the calculation of the Fourier spectrum of the Fourier spectrum. In order to perform an efficient analysis, we compute narrow window filtered Fourier spectra, as proposed by \cite{2006MNRAS.369.1491R}. \cite{mosserapp}, hereafter MA09, have applied and quantified the method, and presented its immediate avantage: the signature of the large separation occurs at a time delay much smaller than the mode lifetime, so that the autocorrelation of the time series is efficient since it stacks cophased signals.

In this work, we address specifically the description of the pipeline based on the autocorrelation of the time series. We recall briefly in Section \ref{EACF} how the envelope autocorrelation function is calculated (the full description is given in MA09). Section \ref{pipeline} presents the main steps of the automated pipeline for the determination of the \emph{mean} large separation, then the determination of the mode envelope. It also shows how to measure the variation of the large separation with frequency. A few characteristics and properties are discussed in Sect. \ref{plus}.

\section{The envelope autocorrelation function\label{EACF}}

The envelope autocorrelation function (EACF) presented in MA09 is first normalized at 1 at null delay, then rescaled to the noise contribution. This rescaling is a crucial step for the automated analysis, since it gives the possibility to test the performance of the method independently of any modeling.
A statistical test as the null hypothesis allows us to define the reliability of the detection.
This rescaling also shows that a tiny peak in the EACF may be in fact fully reliable when compared to the tiniest signature of a white noise.

With this scaling, and taking into account properties of solar-like oscillation spectra, MA09 have shown that an EACF greater than 8 corresponds to the rejection of the null hypothesis at the level 1\,\%. White noise being then unable to explain the autocorrelation signal, one can derive the positive detection of a signal. After correction of possible artefacts, as the signature of the orbital frequency of CoRoT  (\cite{2009arXiv0901.2206A}), this signal can be identified to solar-like oscillations.

\section{Automated pipeline\label{pipeline}}

\subsection{Determination of the mean large separation}

The mean large separation $\dnumoy$ is defined as the mean value of $\deltanunu$, in a frequency range centered on $\numax$ as large as the mode envelope. This asteroseismic parameter is derived first. This means that it is not necessary to first estimate the background due to stellar activity, what makes the asteroseismic analysis direct and simple.

In order to perform efficiently the search, scaling relations are presupposed between $\numax$, $\dnumoy$ and the full-width at half-maximum $\nuenv$ of the mode envelope. The relation between $\numax$ and $\dnumoy$ follows the monomial law reported in CoRoT data (\cite{2009arXiv0906.5002H}, MA09). In order to examine possible outliers, $\pm 30\,\%$ variations around this law are permitted.
We have checked that the relation between $\numax$ and $\dnumoy$ is satisfied for all targets with an agreement better than 20\,\%. The relation between $\nuenv$ and $\dnumoy$ is also derived from CoRoT data (MA09): a ratio $\nuenv / \dnumoy$ of 4 is first supposed for red giants, and 10 for solar-like stars.

The systematic blind analysis consists in searching $\dnumoy$ in a range around an initial guess value $\dnu\ind{g}$. One computes the Fourier spectrum of the windowed Fourier spectrum with a Hanning filter centered on $\numax (\dnu\ind{g})$ and a width $\nuenv(\dnu\ind{g})$. This gives the (possible) signature $\dnu$ close to $\dnu\ind{g}$, if validated by an EACF greater than 8 as presented in Sect.~\ref{EACF}. Large separations in the range from 0.5 to 250\,$\mu$Hz can be exhaustively tested in 18 steps, with the guess values $\dnu\ind{g}$ in progression with a geometric ratio $\sqrt{2}$.

The mean value $\dnumoy$ being firmly established, a more detailed analysis around the interesting frequency range gives then $\numax$ and $\FWHM$. Nevertheless, at this stage, these parameters $\numax$ and $\FWHM$ are less precisely estimated than $\dnumoy$. In fact, the value of $\numax$ does not strictly correspond to the maximum oscillation signal, but to the maximum EACF signal. The difference derives from the fact that the modes lifetime varies with frequency: a mode with low amplitude but large lifetime may have a larger autocorrelation signature than a mode with a larger amplitude and a shorter lifetime.

\subsection{Mode envelope}

The precise determination of $\numax$, $\nuenv$ and of the bolometric amplitude of radial modes can now be performed in a spectrum smoothed with a cosine window of width $\dnumoy$. Then, the background power due to the stellar activity is estimated, in a similar way as performed by other pipelines, in order to estimate the power excess.
We note that the prior determinations of $\numax$ and $\nuenv$ make their refined determination easy.
The bolometric of radial modes is performed according to the recipe described in \cite{2009A&A...495..979M}.

\subsection{Variation of the large separation\label{varidnu}}

If the EACF is high enough, the determination of the variations of the large separation with frequency is possible without any mode fitting, with narrow filters. The examination of the derivative $\diff \dnu / \diff n$ of the large separation with respect to the radial order $n$ is best performed with a filter width approximately equal to $\nuenv / 2$. In order to fully investigate the variation of $\deltanunu$, a filter width equal to $2\,\dnumoy$ is well suited. MA09 have shown that a filter width equal to $0.75\,\dnumoy$ allows the identification of the degree of a ridge, what can be useful for F targets (\cite{2008A&A...488..705A}). The threshold level for analyzing $\deltanunu$ is lower than for the blind analysis, since it benefits from the prior measurement of $\dnumoy$ and from the fact that $\deltanunu$ is now searched in a narrow range: the 1\,\% rejection level is at 4.6 when searching for variation around $\pm 20$\,\% of $\dnumoy$ with a 2-$\dnumoy$ wide filter.

\section{Discussion\label{plus}}

Compared to other pipelines, the pipeline based on the envelope autocorrelation function  shows many qualities:

- It is rapid; the complete operation takes less than 10\,s for the blind analysis of a CoRoT time series with 4\,10$^5$ points, corresponding to a 150-day run at a 32-s sampling time. For the first Kepler 44-day long time series with a sampling about 30\,min, the blind analysis lasts 0.3\,s per target.
Most of these operations are direct calculations or fast Fourier transforms (fft). The use of fft simply requires the scaling presented in Section \ref{EACF} to be performed at the exact frequency resolution.

- It includes the determination of a reliable threshold level. Contrary to other pipelines, the performance and the determination of error bars do not depend on any scaling based on the analysis of theoretical models. Similarly to other pipelines, the EACF just presupposes a solar-like regular pattern.

- It measures the mean large separation with a reliable and reduced error bar. For a solar-like star (resp. a red giant), the determination of $\dnumoy$ at the detection limit is achieved with a relative precision of 0.7\,\% (resp. 1.9\,\%). This precision scales with the inverse of the EACF (values up to 600 have been measured with CoRoT). Since the large separation may significantly vary along the spectrum, the definition and the significance of the mean large may be not accurate enough to fully describe the rich output of this asteroseismic observable (see next point). In the same way, the definition of $\numax$ presupposes a Gaussian mode envelope, what appears to be not valid in all cases, due to the stochastic nature of the mode excitation.

- It gives access to the variation of the large separation without any mode fitting.

\begin{acknowledgements}
This work was supported by the Centre National d'Etudes Spatiales (CNES). It is based on observations with CoRoT.
\end{acknowledgements}


\end{document}